\documentclass[pra,twocolumn,showpacs,groupedaddress,nofootinbib]{revtex4}
\usepackage{amsmath,amssymb,amsfonts}
\usepackage{graphicx}
\usepackage{txfonts}
\newcommand{\ket}[1]{{|#1\rangle}}

\newcommand{\myref}[1]{{(\ref{#1})}}
\newcommand{\D}[1]{{\hat{D}\left( #1 \right)}}
\newcommand{\imwidth}{1.0}

\begin{document}

\title{Quantum resonances in an atom-optical $\delta$-kicked harmonic oscillator}

\author{T. P. Billam}
\affiliation{Department of Physics, Durham University, Durham DH1 3LE, United Kingdom}
\author{S. A. Gardiner}
\affiliation{Department of Physics, Durham University, Durham DH1 3LE, United Kingdom}

\date{\today}

\begin{abstract}
Under certain conditions, the quantum $\delta$-kicked harmonic oscillator displays quantum resonances. We consider an atom-optical realization of the $\delta$-kicked harmonic oscillator, and present a theoretical discussion of the quantum resonances that could be observed in such a system. Having outlined our model of the physical system we derive the values at which quantum resonances occur and relate these to potential experimental parameters. We discuss the observable effects of the quantum resonances, using the results of numerical simulations. We develop a physical explanation for the quantum resonances based on symmetries shared between the classical phase space and the quantum-mechanical time evolution operator. We explore the evolution of coherent states in the system by reformulating the dynamics in terms of a mapping over an infinite, two-dimensional set of coefficients, from which we derive an analytic expression for the evolution of a coherent state at quantum resonance.
\end{abstract}


\maketitle

\section{Introduction}
\label{sec:intro}
As the field of quantum chaos \cite{gutzwiller,haake_book} has emerged over the last few decades, much research has been directed towards the quantal counterparts of classically chaotic Hamiltonian systems \cite{reichl_book,almeida_book}. Amongst these systems, one-dimensional, periodically kicked quantum systems have proved a popular target for study. The dynamics of these systems, which are described by discrete time evolution operators acting over the period of one or more kicks, often display quantum resonances and antiresonances; changes in the degree of localization of the dynamics which are strongly dependent on the value of an effective Planck constant. Advances in the cooling and trapping of atoms and ions has paved the way for substantial direct experimental investigation of such archetypal quantum-chaotic systems, particularly the $\delta$-kicked rotor \cite{casati_etal_1979, izrailev_1979, izrailev_1980, moore_1994, ammann_1998_PRL, dArcy_2001_PRL, moore_1995, dArcy_2003_PRL, ryu_etal_2006, szriftgiser_2002, duffy_etal_2004_PRE}, and also the $\delta$-kicked accelerator \cite{oberthaler_1999, schlunk_2003a, schlunk_2003b,ma_2004, buchleitner_2006, behinaein_2006,fishman_2002,guarneri_rebuzzini_2008}. Another possibility is the $\delta$-kicked harmonic oscillator (KHO) \cite{berman_etal_1991,borgonovi_rebuzzini_1995,carvalho_buchleitner_2004, dana_1994, duffy_etal_2004_PRA, gardiner_etal_1997,gardiner_etal_2000}; many kicked atom (ion) configurations involve trapping potentials which are to some degree harmonic, making the dynamics of the KHO a relevant consideration for a wide variety of experiments.

The classical KHO was originally developed as a model for the interaction of charged particles with electromagnetic fields \cite{zaslavsky_book,zaslavsky_etal_1986,zaslavsky_etal_1988}. The frequency degeneracy of orbits in the harmonic oscillator renders the KAM theorem inapplicable to the KHO (i.e., all invariant tori of the harmonic oscillator may be destroyed at any finite perturbation strength) \cite{arnold_book}. In cases where the kicking period rationally divides the natural oscillator period this allows the KHO to display exotic non-KAM chaos, manifest as interlinked regions of chaotic dynamics forming \textit{stochastic web\/} structures throughout phase space \cite{zaslavsky_book,hu_1999}. The quantum KHO has also been studied in considerable depth \cite{dana_1994,dana_1995_phys_lett_a,dana_1995_PRE,dana_dorofeev_2005_PRE, shepelyansky_sire_1992}; particularly in the case where it is kicked four times per oscillator period, when it can be related to the kicked Harper model \cite{artuso_etal_1992,artuso_etal_1992_b,borgonovi_shepelyansky_1995,guarneri_1989,guarneri_1993,guarneri_borgonovi_1993}. Depending on the relative alignment of the kicking potential with respect to the trapping potential, the KHO can display either quantum resonances (even kicking potentials) or quantum antiresonances (odd kicking potentials) \cite{dana_dorofeev_2005_PRE}. 

In this paper, we consider the effects of quantum resonances that may be observed in an atom-optical KHO. In Section \ref{sec:aokho} we outline our model of the atom-optical KHO and introduce the kick-to-kick Floquet operator. In Section \ref{sec:origin} we derive the structure of the quantum resonances and their effects on the dynamics. We consider the relation between the quantum resonances, the eigenstates of the Floquet operator for $q$ kicks, and that Floquet operator's expansion in terms of phase space displacement operators; this leads to an alternative, physically intuitive explanation for the occurrence of quantum resonances.  In Section \ref{sec:manifestation} we consider the measurable manifestation of quantum resonant dynamics in the atom-optical KHO, including the experimental accessibility of the necessary parameter regimes.  Section \ref{sec:conclusions} consists of the conclusions, which are then followed by three technical appendices.

\section{Atom-optical $\delta$-kicked harmonic oscillator}
\label{sec:aokho}

\subsection{System Hamiltonian}
We consider a two-level atom (or ensemble of noninteracting atoms) or ion, subject to tight radial confinement and relatively loose harmonic axial confinement, and periodically driven by pulses from an off-resonant laser standing wave. We assume the pulse duration to be sufficiently short for the particle centre of mass displacement during the pulse to be negligible relative to the period of the standing wave (Raman-Nath regime).  To a good approximation, the excited internal state can be eliminated, and the system can be described by the one-dimensional single particle Hamiltonian \cite{gardiner_etal_1997,gardiner_etal_2000,saunders_etal_2007}
\begin{equation}
 \hat{H} = \frac{\hat{p}^2}{2M} + \frac{M \omega^2 \hat{x}^2}{2} + \frac{\hbar\Omega^{2}t_{p}}{8\Delta}\cos(K \hat{x}) \sum_{n=-\infty}^{\infty} \delta(t - nT),
 \label{eqn:quantum_hamiltonian}
\end{equation}
describing the centre of mass motion in the axial direction only.  Here $M$ is the particle mass, and $\omega$ the frequency of the harmonic trapping potential; the kicks are pulses of duration $t_{p}$, repeated with period $T$, from the laser standing wave, detuned by $\Delta$ from the two-level transition and with Rabi frequency $\Omega$, where $K=4\pi/\lambda_{L}$ and $\lambda_{L}$ is the laser wavelength.\footnote{The quantity $\hbar K$ is therefore equal to two photon recoils.}

We have set the origin of the $x$-coordinate to be at the minimum of the harmonic potential, and choose the laser standing wave to be aligned symmetrically about this minimum. The resulting cosine potential is an even function in Hamiltonian  \myref{eqn:quantum_hamiltonian}; hence, we can expect quantum resonances, i.e., particular parameter regimes associated with ballistic growth in the system energy, to be observable \cite{dana_1994,dana_dorofeev_2005_PRE}.  Shifting the standing wave by $\lambda_{L}/8$ yields a sine potential --- an odd function, producing antiresonances in  parameter regimes where quantum resonances occur for a cosine potential.  When the alignment of the laser standing wave with respect to the trapping minimum lies between these extremes, intermediate quantum resonant behaviour can occur \cite{dana_1994,dana_dorofeev_2005_PRE}.  It is therefore not essential for the laser standing wave to be aligned perfectly symmetrically in order for quantum resonant behaviour to be observable, although from now on we assume this, for simplicity. More important is that any kick-to-kick variations in the standing wave alignment be negligible compared to the standing wave period, otherwise the resulting randomness introduced into the dynamics will destroy the desired quantum resonant behaviour.

\subsection{Time evolution}
The time evolution of the quantum KHO is described by a Floquet (discrete time evolution) operator $\hat{F}$ which maps the state of the system at a time immediately prior to one kick onto the state at a time immediately prior to the next.  It is convenient to describe the Floquet operator $\hat{F}$ in terms of the  harmonic oscillator creation and annihilation operators $\hat{a}^\dagger$, $\hat{a}$, where $\hat{a}=\sqrt{m\omega / 2\hbar}(\hat{x}+i\hat{p}/m\omega)$.  Hence,
\begin{equation}
 \hat{F} = e^{-i(\hat{a}^\dagger \hat{a} + 1/2)\tau} e^{-i\kappa\mathrm{cos}[\eta(\hat{a} + \hat{a}^\dagger)]/\sqrt{2}\eta^2},
\label{eqn:floquet_operator}
\end{equation}
where $\tau = \omega T$ is the dimensionless kicking period, $\kappa=\hbar\Omega^{2}t_{p} K^2/8\sqrt{2}\Delta M\omega$ is the dimensionless kicking strength, and  $\eta = K \sqrt{\hbar / 2M\omega}$ is the  Lamb-Dicke parameter, which characterizes the scale of the harmonic ground state relative to the laser wavelength. The parameters $\tau$ and $\kappa$ emerge naturally from a rescaling to dimensionless dynamical variables of the KHO classical map  \cite{zaslavsky_book,zaslavsky_etal_1986,zaslavsky_etal_1988}, whereas $\eta^{2}$ now assumes the role of an effective Planck constant \cite{gardiner_etal_1997}.

\subsection{Phase space symmetries}
\label{subsec:crystal_quasicrystal_symmetry}
We confine our attention to when the kicking period rationally divides the natural period of the oscillator, i.e., when $\tau = 2\pi r/q$ for coprime natural numbers $r$, $q$.  Except in the special cases where $q=1$ or $2$, the phase space of the classical KHO is then spanned by a \textit{stochastic web\/} \cite{zaslavsky_book,hu_1999},
i.e., interconnected regions of chaotic dynamics penetrating through all of phase space. If $q \in q_c \equiv \{1,2,3,4,6\}$ the classical phase space has a discrete translational (crystal) symmetry; otherwise the stochastic web has a quasicrystal structure.  These crystal symmetries are reflected in the quantum regime through the existence of infinite sets of
unitary displacement operators\footnote{Here $\alpha$ is the conventional complex index of a coherent state, such that $\hat{D}(\alpha)\vert0\rangle = \vert \alpha \rangle$ and $\hat{a}\vert\alpha\rangle = \alpha\vert\alpha\rangle$.} $\hat{D}(\alpha) = e^{\alpha \hat{a}^\dagger - \alpha^\ast \hat{a}}$ commuting with $\hat{F}^q$, the evolution operator for one natural oscillator period \cite{berman_etal_1991,borgonovi_rebuzzini_1995,gardiner_etal_1997,gardiner_etal_2000}.  This implies that eigenstates of $\hat{F}^q$ for $q\in q_{c}$ have a comparable crystal structure, and, importantly, are consequently \textit{extended} in phase space.  In the non-trivial crystal cases $q\in\{3,4,6\}$ --- when the displacement operators form a two-dimensional crystal lattice in phase space --- the set of displacements commuting with $\hat{F}^q$ is $\{\hat{D}(\gamma_q)\}$, where
\begin{align}
\gamma_4     &= k\pi/\eta + il\pi/\eta, \label{eqn:classical_q=4_group} \\
\gamma_3 = \gamma_6 &= k \left( 1+i/\sqrt{3} \right) \pi/\eta + l \left( 1-i/\sqrt{3} \right) \pi/\eta, \label{eqn:classical_q=36_group}
\end{align}
for arbitrary integer $k$, $l$.\footnote{To reinforce the connection to the classical stochastic web, one can rescale the classical realization in terms of an appropriate dimensionless position -- momentum pair and obtain (for the cases $q=3,4,6$) the translations \myref{eqn:classical_q=4_group} and \myref{eqn:classical_q=36_group} with $\eta$ set to unity \cite{gardiner_etal_2000}.}

\section{Origin of quantum resonances}
\label{sec:origin}

\subsection{Quantum resonant values of $\eta^{2}$}
\label{subsec:occurrence}
The general study of quantum resonances in systems with periodic classical phase spaces has been carried out in considerable theoretical detail by Dana \cite{dana_1994,dana_1995_phys_lett_a,dana_1995_PRE}. Our intention here is to provide an accessible derivation of the conditions under which quantum resonances occur specifically for the atom-optical KHO, as described in Sec.\ \ref{sec:aokho}.  

With this in mind, we substitute $\tau = 2\pi r/q$ into \myref{eqn:floquet_operator}, and rearrange  $\hat{F}^q$ into the product
\begin{equation}
 \hat{F}^q = \prod_{j=0}^{q-1} e^{i(\hat{a}^\dagger \hat{a} + 1/2)2\pi jr/q} e^{-i\kappa\mathrm{cos}[\eta(\hat{a} + \hat{a}^\dagger)]/\sqrt{2}\eta^2} e^{-i(\hat{a}^\dagger \hat{a} + 1/2)2\pi jr/q}.
\label{eqn:product_rearrangement}
\end{equation}
We use the identity $e^{i\hat{a}^\dagger \hat{a}\tau}(\hat{a} + \hat{a}^{\dagger}) e^{-i\hat{a}^\dagger \hat{a}\tau} = \hat{a}e^{-i\tau} + \hat{a}^{\dagger}e^{i\tau}$ to transform each term of \myref{eqn:product_rearrangement}, and write $\hat{F}^q$ as
\begin{equation}
\hat{F}^q = \prod_{j=0}^{q-1} \exp \left(-i\kappa\cos\left[\eta\left(\hat{a} e^{-i2\pi jr/q} + \hat{a}^\dagger e^{i2\pi jr/q} \right)\right] / \sqrt{2} \eta^2 \right),
\label{eqn:q-kicks-q-axes}
\end{equation}
that is, as a sequence of $q$ cosine kicks directed along $q$ axes equally distributed in angle about the origin in phase space.  

Using the identity \cite{gradshteyn} 
\begin{equation}
e^{i\chi\cos(\xi)} = \sum_{j=-\infty}^{\infty}i^j J_j(\chi) e^{ij\xi}
\label{eqn:jacobi_anger}
\end{equation} 
(the $J_{j}$ are $j$th-order cylindrical Bessel functions), we now expand $\hat{F}^q$ in terms of products of unitary displacement operators:
\begin{equation}
\hat{F}^q =  \prod_{j=0}^{q-1}   \sum_{k_j = -\infty}^{\infty}  i^{k_j}  J_{k_j}\left(-\kappa/\sqrt{2}\eta^2 \right) \hat{D} \left( i k_j \eta e^{i2\pi jr/q}  \right).
\label{eqn:ultimate_displacement_expansion}
\end{equation}
Taking the commutation relation for the unitary displacement operators $\hat{D}(\alpha)\hat{D}(\beta) = e^{\alpha \beta^\ast - \alpha^\ast \beta}\hat{D}(\beta)\hat{D}(\alpha)$ \cite{gardiner_zoller}, it follows that for $\alpha =i k_m \eta e^{i2\pi mr/q}$, $\beta = i k_n \eta e^{i2\pi nr/q}$, then $e^{\alpha \beta^\ast - \alpha^\ast \beta} = 1$ (i.e., the displacement operators commute) whenever
\begin{equation}
2\eta^2 k_m k_n \sin \left( \frac{2\pi r[m-n]}{q} \right) = 2\pi w,
\label{eqn:master_resonance_condition}
\end{equation}
for integer $w$.  If \myref{eqn:master_resonance_condition} is fulfilled for all $m-n \in \{0,1,\ldots,q-1\}$ it follows that all displacement operators in \myref{eqn:ultimate_displacement_expansion} mutually commute, and hence all of the product terms in \myref{eqn:q-kicks-q-axes} must mutually commute.  Hence, under these conditions, the $v$th power of $\hat{F}^q$ will describe a $q$-fold diffraction exactly equivalent to \myref{eqn:q-kicks-q-axes} but with amplified kicking strength $\kappa v$:
\begin{equation}
\hat{F}^{qv} = \prod_{j=0}^{q-1} \exp \left(-i\kappa v\cos\left[\eta \left(\hat{a} e^{-i2\pi jr/q} + \hat{a}^\dagger e^{i2\pi jr/q} \right) \right] / \sqrt{2} \eta^2 \right).
\label{eqn:many-kick-reduction}
\end{equation}
This is analogous to the case where the quantum resonance condition is fulfilled for the quantum $\delta$-kicked rotor, when the free evolution operator between kicks collapses to the identity, and the evolution induced by $v$ iterations of the Floquet operator is equivalent to that caused by a single kick with $v$ times the kicking strength \cite{casati_etal_1979,moore_1995,saunders_etal_2007}.

The values of $\eta^{2}$ for which \myref{eqn:master_resonance_condition} is fulfilled therefore determine when quantum resonances occur in the KHO.  In summary, \myref{eqn:master_resonance_condition} is fulfilled for: all values of $\eta^2$ when $q\in\{1,2\}$; $\eta^2$ equal to a multiple of $\pi$ when $q=4$; and  $\eta^2$ equal to a multiple of $2\pi/\sqrt{3}$ when $q\in \{3,6\}$ (see Appendix \ref{app:derivation}). For $q\notin q_c$ it is impossible to satisfy \myref{eqn:master_resonance_condition}.  Note that, for $q\in\{1,2\}$, \myref{eqn:q-kicks-q-axes} reduces to being exactly equivalent to what one would expect for the quantum $\delta$-kicked rotor fulfilling the conditions necessary for quantum resonance; the free evolution between kicks has no net effect, and the (time-periodic) dynamics can be equivalently described by a single amplified kick.  However, the fact that the free evolution between kicks has no net effect is equally true classically (for $q=1$ this simply means that after one oscillator period between kicks, the oscillator has returned to its original location
in phase space), and one must therefore conclude that quantum resonances do not occur for $q\in\{1,2\}$, as is consistent with the dynamics' independence of the effective Planck constant $\eta^{2}$.

Consequently the KHO displays quantum resonances only if $q \in \{3,4,6\}$, and for $\eta^2$ equal to multiples of the principal quantum resonance values: $\pi$ when $q=4$, and $2\pi/\sqrt{3}$ when $q=3$ or $q=6$.  At further rational multiples $a/b$ of the principal quantum resonance values, commutativity between kicks depends on the value of $k_m k_n$ in \myref{eqn:master_resonance_condition}, and only occurs for some terms within the summation in \myref{eqn:ultimate_displacement_expansion}, generating higher order quantum resonances. The proportion of terms for which the displacements commute decreases with the denominator $b$, approaching zero as $a/b$ approaches irrationality.

\subsection{Relation to phase space symmetries of $\hat{F}^q$}
When $q \in \{3,4,6\}$, the set of unitary displacements in \myref{eqn:ultimate_displacement_expansion} describe a two-dimensional crystal lattice in phase space. In analogy with Section \ref{subsec:crystal_quasicrystal_symmetry} we denote this set $\{\hat{D}(\Gamma_q)\}$, where
\begin{align}
 \Gamma_4     &= k \eta + il \eta, \label{eqn:quantum_q=4_set} \\
 \Gamma_3 = \Gamma_6 &= k \left( \frac{\sqrt{3}+i}{2} \right) \eta + l \left( \frac{\sqrt{3}-i}{2} \right) \eta. \label{eqn:quantum_q=36_set}
\end{align}
Comparing \myref{eqn:quantum_q=4_set} and \myref{eqn:quantum_q=36_set} with \myref{eqn:classical_q=4_group} and \myref{eqn:classical_q=36_group}, one sees that, at the \textit{principal quantum resonance values\/} [$\eta^2 =\pi$ ($q=4$), or $\eta^2 = 2\pi/\sqrt{3}$ ($q=3,6$)], the set  $\{\hat{D}(\Gamma_q)\}$ of displacements constituting $\hat{F}^q$ is identical to the set $\{\hat{D}(\gamma_q)\}$ of translational symmetries of the eigenstates of $\hat{F}^q$.  Moreover, for $\eta^2$ equal to rational multiples $a/b$ of these principal values, $\hat{D}(\Gamma_q(k,l))$ remains an element of $\{\hat{D}(\gamma_q)\}$ whenever $k$ and  $l$ are integer multiples of $b$.  This (complete or partial) equivalence of $\{\hat{D}(\Gamma_q)\}$ and $\{\hat{D}(\gamma_q)\}$ offers a physical explanation for quantum resonances in the KHO; at quantum resonance values of $\eta^2$ the displacements constituting each kick are translational symmetries of $\hat{F}^q$'s eigenstates, and consequently commute with $\hat{F}^q$.

\subsection{Evolution of an initial coherent state}
\label{subsec:evolution}

\subsubsection{Motivation}
In the context of cold, trapped atoms or ions, the initial motional state of the atom(s) or ion in any experiment will generally be as close to a zero-temperature state as is practicable for the specific experimental configuration.  This provides a clear motivation to study the system evolution when the initial condition is a harmonic oscillator ground state. In the case of an initial atomic Bose-Einstein condensate, this would describe the initial condensate mode if atom-atom interactions could be neglected, which could potentially be achieved with the aid of magnetic fields to exploit a Feshbach resonance \cite{gustavsson_etal_2008}.  

Coherent states are harmonic oscillator ground states which have been displaced in phase space, and as such can in principle be accessed by displacing an initial ground state in momentum (e.g., by application of a Bragg pulse) or position (e.g., by suddenly changing the trap minimum), possibly in combination with a free evolution subsequent to the displacement.  In what follows we therefore consider the evolution of an initial coherent state analytically, which,  in the case of quantum resonance, proves to be comparatively tractable.

\subsubsection{Phase space lattice state expansion}
We begin by applying the identity \myref{eqn:jacobi_anger} to the Floquet operator \myref{eqn:floquet_operator}, yielding
\begin{equation}
\label{eqn:action_on_expansion_1}
 \hat{F} = e^{-i(\hat{a}^\dagger \hat{a} + 1/2) 2\pi r/q} \sum_{k=-\infty}^{\infty} i^{k} J_{k}(\zeta) \D{ i \eta k},
\end{equation}
where we have introduced $\zeta = -\kappa / \sqrt{2}\eta^{2} = -\Omega^{2} t_{p}/8 \Delta$ \footnote{$\zeta$ will usually be a positive quantity, as this corresponds to red detuning.}. It follows that applying $\hat{F}$ to an arbitrary initial coherent state $\ket{\alpha}$ will yield a weighted superposition of regularly spaced coherent states.  More generally, we will show that if the state of the system $|\psi_{j}\rangle$ after $j$ applications of \myref{eqn:action_on_expansion_1} takes the form of a ``lattice state''
\begin{equation}
 \label{eqn:lattice_expansion_def}
\ket{\psi_j} = \sum_{m,n = -\infty}^{\infty} M^{[j]}_{m,n} \D{ i\eta \left[m + n e^{-i2\pi r/q} \right] } e^{-i\pi jr/q} \ket{\alpha e^{-i2\pi jr/q}},
\end{equation}
then $\ket{\psi_{j+1}} = \hat{F}\ket{\psi_{j}}$ will take an essentially similar form.  The $\ket{\alpha e^{-i2\pi jr/q}}$ in \myref{eqn:lattice_expansion_def} are all coherent states, and the $M^{[j]}_{m,n}$ a set of complex coefficients.  

Applying \myref{eqn:action_on_expansion_1} to \myref{eqn:lattice_expansion_def}, and using the identities \cite{gardiner_zoller}
 $\D{\alpha} \D{\beta} = e^{(\alpha \beta^\ast - \alpha^\ast \beta)/2} \D{\alpha + \beta}$
and
$ e^{-i\hat{a}^\dagger \hat{a} \tau} \D{\alpha} \ket{\beta} =
\hat{D}(\alpha e^{-i\tau})\ket{\beta e^{-i\tau}}$,
now yields, in the first instance,
\begin{equation}
 \label{eqn:action_on_expansion_2}
\begin{split}
 \hat{F} \ket{\psi_{j}} =& \sum_{k,m,n = -\infty}^{\infty}  i^{k} J_{k}(\zeta) M^{[j]}_{m,n} e^{i k n \eta^2 \sin(2\pi r/q)} \\ 
& \times \D{ i\eta \left[(m+k) e^{-i2\pi r/q} + n e^{-i4\pi r/q} \right] } \\
& \times e^{-i(j+1)r/q} \ket{\alpha e^{-i2\pi(j+1)r/q}}.
\end{split}
\end{equation}
When $q\in \{3,4,6\}$, $e^{-i4\pi r/q}$ can be written as an integer superposition of unity and $e^{-i2\pi r/q}$, allowing  \myref{eqn:action_on_expansion_2} to be simplified. From now on we assume $r=1$ to illustrate the procedure:
\begin{equation}
\label{eqn:vector_relations}
 e^{-i4\pi /q} = \left\{
\begin{array}{l l}
  -1-e^{-i2\pi /q} & \quad \mbox{if $q=3$},\\
  -1 & \quad \mbox{if $q=4$},\\
  -1+e^{-i2\pi /q} & \quad \mbox{if $q=6$}. \end{array}  \right.
\end{equation}
Using \myref{eqn:vector_relations} to make replacements in \myref{eqn:action_on_expansion_2}, one obtains
\begin{equation}
 \label{eqn:action_on_expansion_3}
\begin{split}
 \hat{F} \ket{\psi_{j}} = & \sum_{k,m,n = -\infty}^{\infty} i^{k} J_{k}(\zeta) M^{[j]}_{m,n} e^{i k n \eta^2 \sin(2\pi /q)} \\
& \times \D{ i\eta \left[-n + (m+n\xi_{q}+k) e^{-i2\pi /q} \right] } \\
& \times e^{-i(j+1)/q} \ket{\alpha e^{-i2\pi(j+1)/q}},
\end{split}
\end{equation}
where we have introduced
\begin{equation}
 \xi_{q} = \left\{
\begin{array}{l l}
  -1 & \quad \mbox{if $q=3$,}\\
  0 & \quad \mbox{if $q=4$,}\\
  1 & \quad \mbox{if $q=6$.} \end{array}  \right.
\end{equation}
Effecting a change of indices such that $m \rightarrow m^\prime=-n$ and $n \rightarrow n^\prime=m+n\xi_{q}+k$ now gives
\begin{equation}
 \label{eqn:action_on_expansion_4}
\begin{split}
 \hat{F} \ket{\psi_{j}} = & \sum_{k,m,n = -\infty}^{\infty} i^{k} J_{k}(\zeta) M^{[j]}_{m\xi_{q}+n-k,-m} e^{-i k m \eta^2 \sin(2\pi /q)} \\
& \times \D{ i\eta \left[m + n e^{-i2\pi /q} \right] }e^{-i(j+1)/q} \ket{\alpha e^{-i2\pi(j+1)/q}},
\end{split}
\end{equation}
where we have dropped the primes, for convenience.  With \myref{eqn:action_on_expansion_4} we have an expression of the same form as \myref{eqn:lattice_expansion_def}, but for the state $\ket{\psi_{j+1}}$, and can obtain the evolution resulting from $\hat{F}$ as a mapping of the lattice coefficients:
\begin{equation}
\label{eqn:coefficient_mapping}
M^{[j+1]}_{m,n} = \sum_{k=-\infty}^{\infty} i^{k} J_{k}(\zeta) M^{[j]}_{m\xi_{q}+n-k,-m} e^{-i k m \eta^2 \sin(2\pi /q)}.
\end{equation}

The coefficient mapping \myref{eqn:coefficient_mapping} fully describes the time evolution of lattice states in general. The coherent state at the center of the lattice state when $j = 0\mod q$ can be chosen arbitrarily. Hence, choosing an initial set of coefficients
\begin{equation}
 M^{[0]}_{m,n} = \delta_{m,0} \delta_{n,0},
\label{eqn:single_coherent_state}
\end{equation}
amounts to following the evolution of an initial coherent state $\ket{\alpha}$. 

The general evolution of the coefficients over many kicks described by \myref{eqn:coefficient_mapping} yields an ever-lengthening product of nested sums that is not especially insightful.  In the case of quantum resonance, however, $\eta^2 \sin(2\pi /q) = w\pi$, which simplifies the final exponent in \myref{eqn:coefficient_mapping}, allowing a simpler expression for evolution over many kicks to be extracted. In the next section we follow the evolution of an initial coherent state, for $q=4$ and $\eta^2 = \pi$, by repeatedly applying the mapping \myref{eqn:coefficient_mapping}.

\subsubsection{Analytic solution at quantum resonance when $q=4$}
\label{subsubsec:q4}
Our initial condition is a single coherent state, as described by  \myref{eqn:single_coherent_state}.  We first rearrange the indices to obtain
$ M^{[0]}_{n-k,-m} = \delta_{n-k,0} \delta_{-m,0} = \delta_{n,k} \delta_{m,0}$,
which we substitute into \myref{eqn:coefficient_mapping}, yielding
\begin{equation}
 M^{[1]}_{m,n} = \sum_{k=-\infty}^{\infty} i^{k} J_{k}(\zeta) \delta_{n,k} \delta_{m,0} e^{-ikm\pi}
= i^{n} J_{n}(\zeta) \delta_{m,0} e^{-imn\pi}.
\end{equation}
We now repeat this procedure; rearranging the indices and using the identity $i^n J_n(\zeta) = i^{-n} J_{-n}(\zeta)$ yields
\begin{equation}
 M^{[1]}_{n-k,-m}= i^{m} J_{m}(\zeta) \delta_{n,k} e^{i(n-k)m\pi},
\end{equation}
which we substitute into \myref{eqn:coefficient_mapping}:
\begin{align}
M^{[2]}_{m,n}
&= i^{m} J_{m}(\zeta) i^{n} J_{n}(\zeta)  e^{-imn\pi}, \label{eqn:M2_final}\\
\Rightarrow M^{[2]}_{n-k,-m} &= i^{n-k} J_{n-k}(\zeta)  i^{m} J_{m}(\zeta)e^{i(n-k)m\pi}. \label{eqn:M2}
\end{align}
We next substitute \myref{eqn:M2} into \myref{eqn:coefficient_mapping}, accompanied by a change of indices $k \rightarrow -k$ and use of $i^k J_k(\zeta) = i^{-k} J_{-k}(\zeta)$:
\begin{equation}
\begin{split}
M^{[3]}_{m,n}
&   =i^{m} J_{m}(\zeta) i^n  e^{imn\pi} \sum_{k=-\infty}^{\infty}  J_{k}(\zeta)   J_{n+k}(\zeta) e^{ik\pi} .
\label{eqn:pre_graf}
\end{split}
\end{equation}
The summation over $k$ can be reduced using Graf's addition theorem (Appendix \ref{app:graf}) to give
\begin{equation}
 M^{[3]}_{m,n} = i^{m} J_{m}(\zeta) i^{n} J_{n}(2\zeta) e^{imn\pi}.
\end{equation}
We repeat the process, again with a change of indices $k \rightarrow -k$ followed by the application of Graf's theorem:
\begin{align}
 M^{[3]}_{n-k,-m} &= i^{n-k} J_{n-k}(\zeta) i^{m} J_{m}(2\zeta) e^{-i(n-k)m\pi},\\
\begin{split}
\Rightarrow M^{[4]}_{m,n} 
&= i^{m} J_{m}(2\zeta) i^n e^{-imn\pi} \sum_{k=-\infty}^{\infty} J_{k}(\zeta) J_{n+k}(\zeta) e^{ik\pi} \\
&=  i^{m} J_{m}(2\zeta) i^{n} J_{n}(2\zeta)  e^{-imn\pi}. 
\end{split}
\label{eqn:M4_final}
\end{align}

\begin{figure}
\includegraphics[width=\imwidth\columnwidth]{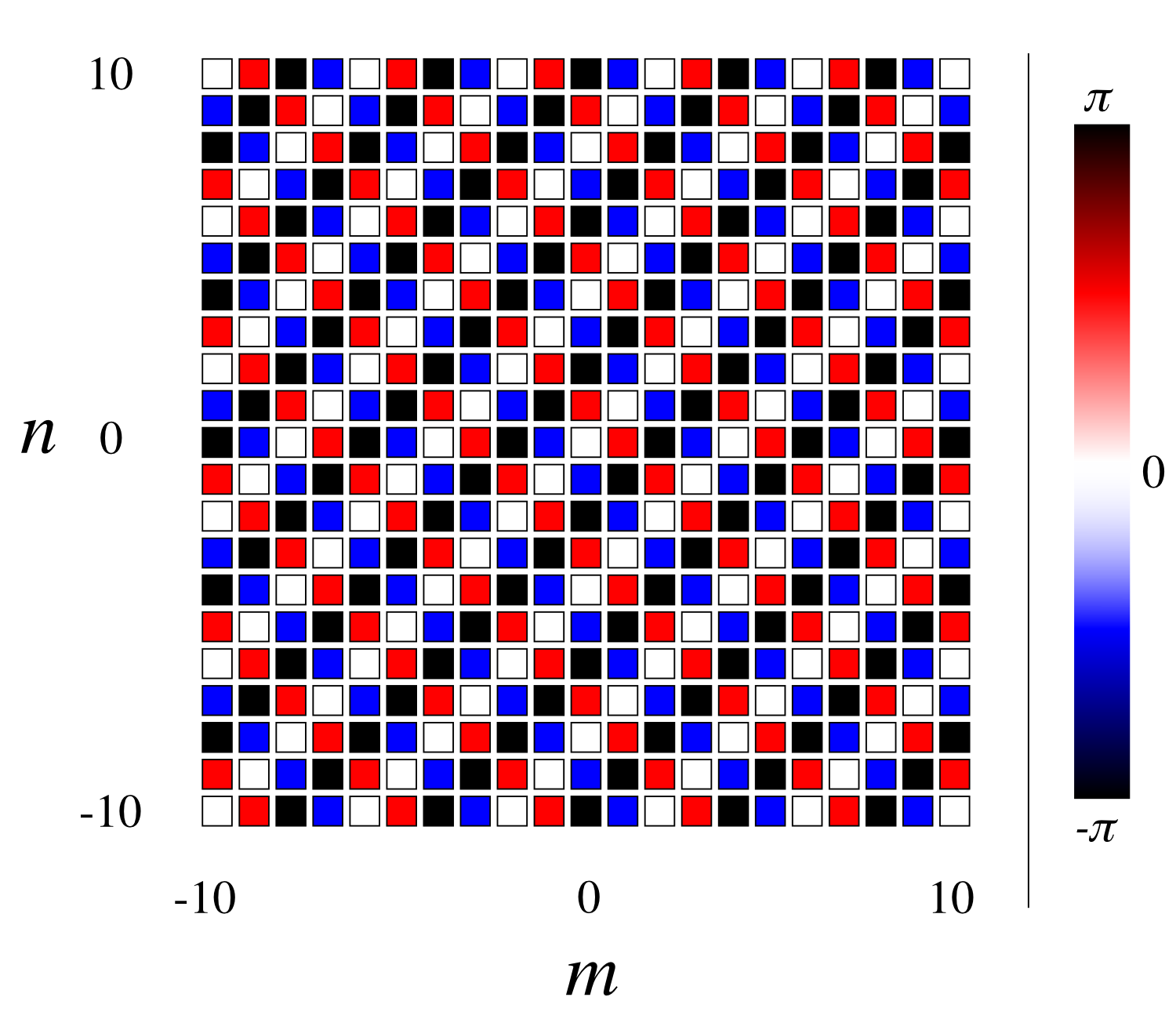}
\caption{(color online). Phases of the coefficients $M^{[N]}_{m,n}$  as given by \myref{eqn:final_result}, which describe the state of a KHO (with $q=4$) after the application of $N$ ($N \geq2$) kicks to an initial coherent state, for the quantum resonant case $\eta^2 = \pi$. The phases shown neglect the parity of the Bessel functions in the coefficient [that is, we actually plot $M^{[N]}_{m,n} / J_{m}(C_m^{[N]}\zeta) J_{n}(C_n^{[N]}\zeta) = (-1)^{mn}i^{m+n}$] so that the pattern remains constant for all $N \geq 2$.}
\label{fig:lattice_phases}
\end{figure}

The final expression obtained for $M^{[4]}_{m,n}$ \myref{eqn:M4_final} is identical to that for $M^{[2]}_{m,n}$ \myref{eqn:M2_final} but with $\zeta$ consistently replaced by $2\zeta$. Further evolution will see such cycles repeated; the time evolution of the coefficients is entirely contained within the changing arguments of the Bessel functions:
\begin{equation}
 M^{[N]}_{m,n} = (-1)^{mn} i^m J_{m}(C_m^{[N]}\zeta)i^n  J_{n}(C_n^{[N]}\zeta) ,
 \label{eqn:final_result}
\end{equation}
where 
\begin{align}
 C_m^{[N]} &= \left\{
\begin{array}{l l}
  2^{(N-2)/2} & \quad \mbox{$N$ even}\\
  2^{(N-3)/2} & \quad \mbox{$N$ odd} \end{array},  \right. \\
 C_n^{[N]} &= \left\{
\begin{array}{l l}
  2^{(N-2)/2} & \quad \mbox{$N$ even}\\
  2^{(N-1)/2} & \quad \mbox{$N$ odd} \end{array}.  \right.
\end{align}
As shown in Fig.\ \ref{fig:lattice_phases}, in the case of quantum resonance the phases of the lattice state coefficients [Eq.\ \myref{eqn:final_result}] take an extremely simple form, which repeats regularly across all of phase space.  We emphasize that this holds for any initial coherent state.  

We have chosen the case $\eta^2=\pi$ for simplicity, however the above analysis holds for all $\eta^2 = w\pi$. For more general $\eta$, each application of Graf's theorem introduces a $k$-dependent phase (see Appendix \ref{app:graf}), preventing an end result as straightforward as \myref{eqn:final_result}.  As a consequence, no such regularity of the phases, as shown in Fig.\ \ref{fig:lattice_phases}, is apparent for non-resonant values of $\eta^{2}$ beyond the first two kicks.

\subsubsection{Analytic solution at quantum resonance when $q=3,6$}
In these cases the locations in phase space of the coherent states making up the lattice states \myref{eqn:lattice_expansion_def}  are defined in terms of $i$ and $ie^{-i2\pi/q}$, which represent two non-orthogonal directions in the complex plane. This non-orthogonality is necessary to derive the coefficient mapping \myref{eqn:coefficient_mapping}, but it adds complexity to the evolution of the coefficients for a coherent state initial condition. In contrast to the case $q=4$, where all coefficients $M^{[j]}_{m,n}$ are a direct function of $m$ and $n$ at quantum resonance, in the cases $q=3$ and $q=6$ the coefficients $M^{[j]}_{m,n}$ for $j \geq 3$ can only be expressed as a summation over products of three Bessel functions, even at quantum resonance.

Despite this added complexity, the cases $q=3$ and $q=6$ remain tractable at quantum resonance. There is a three-step cycle in the coefficients, analogous to the two-step cycle in the coefficients occurring when $q=4$, with the coefficients $M^{[3(j+1)]}_{m,n}$ being equal to $M^{[3j]}_{m,n}$ but with the arguments of all Bessel functions doubled. A derivation for the case $q=6$ is given in Appendix \ref{app:evo}.

\section{Manifestation of quantum resonances}
\label{sec:manifestation}
\subsection{Ballistic phase space expansion}
\begin{figure}
\includegraphics[width=\imwidth\columnwidth]{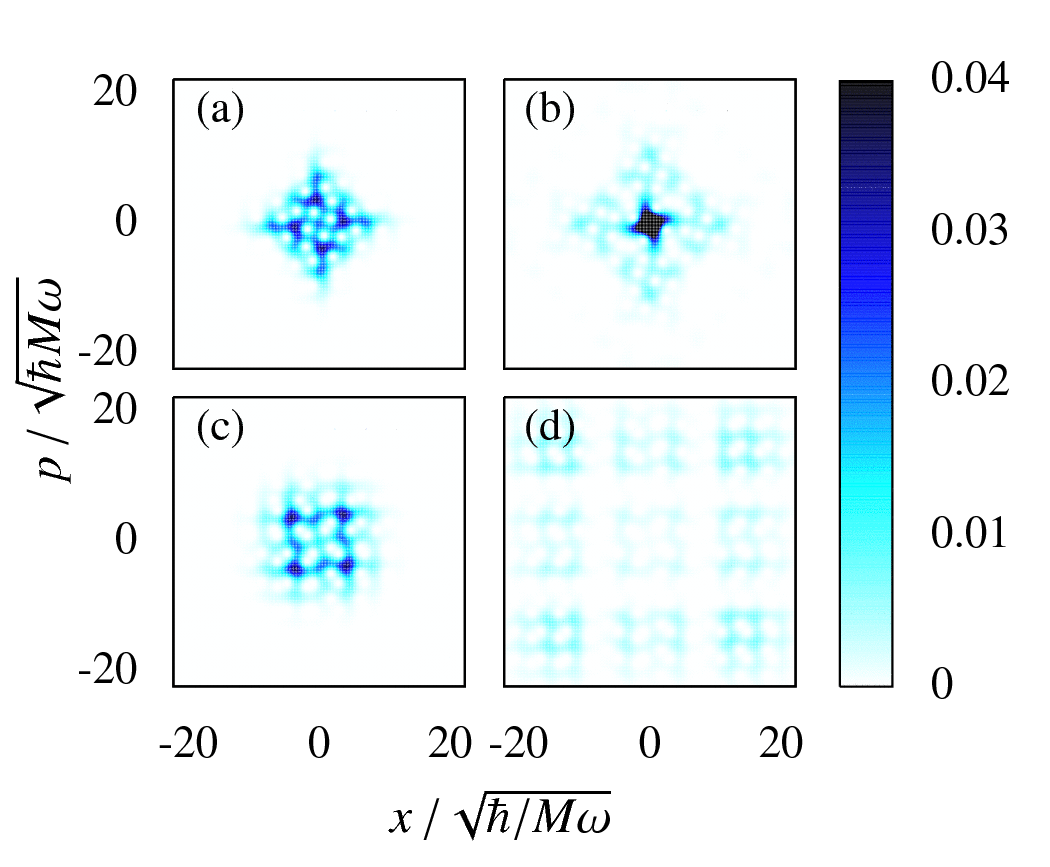}
\caption{(color online). $Q$ functions $Q_{j}(\alpha,\alpha^{*})=\vert \langle \psi_{j} \vert \alpha \rangle \vert ^2 / \pi$ \cite{gardiner_zoller} for the KHO (with $q=4$ and $\kappa = -0.8$) evolved from the harmonic oscillator ground state. 
We contrast the non-resonant case, where $\eta^{2} = \phi\pi$ [$\phi = (1+\sqrt{5})/2$ is the golden ratio] for (a) $j=36$ kicks and (b) $j=108$ kicks, with the resonant case $\eta^{2} = \pi$ for (c) $j=36$ and (d) $j=108$.  Although $Q_{j}(\alpha,\alpha^{*}) > 0.04$ close to the origin in (b), our scale is fixed at this maximum to aid comparison between the images.}
\label{fig:q-functions}
\end{figure}

As for the atom-optical $\delta$-kicked rotor \cite{moore_1994,saunders_etal_2007} or $\delta$-kicked accelerator \cite{saunders_etal_2009,halkyard_etal_2008} systems, the primary observable effect of the quantum resonances in the KHO is an enhanced absorption of energy from the kicking potential.  This is visible as an enhanced rate of expansion of the oscillator's state through phase space, as can be seen in Fig.\ \ref{fig:q-functions}, which shows $Q$ functions (Husimi distributions)  \cite{gardiner_zoller} $Q_{j}(\alpha,\alpha^{*}) = |\langle \psi_{j}|\alpha\rangle |^{2}/\pi$ evolved from an initial harmonic oscillator ground state under resonant and non-resonant conditions.  The evident symmetry between the position and momentum quadratures of the $Q$ functions in Fig.\ \ref{fig:q-functions} mean \textit{in situ\/} measurement of the position density or time-of-flight determination of the momentum density would both provide equally effective signatures of quantum resonant evolution, and the $Q$ function itself may also in principle be determined directly \cite{gardiner_etal_1997,lutterbach_1997}. Note, however, that in any experiment the harmonic potential cannot be infinite in extent; after a large number of kicks the state of the oscillator will extend beyond the harmonic region. Consequently, the subsequent time evolution will depart from the predictions of our, entirely harmonic, model.

\begin{figure}
\includegraphics[width=\imwidth\columnwidth]{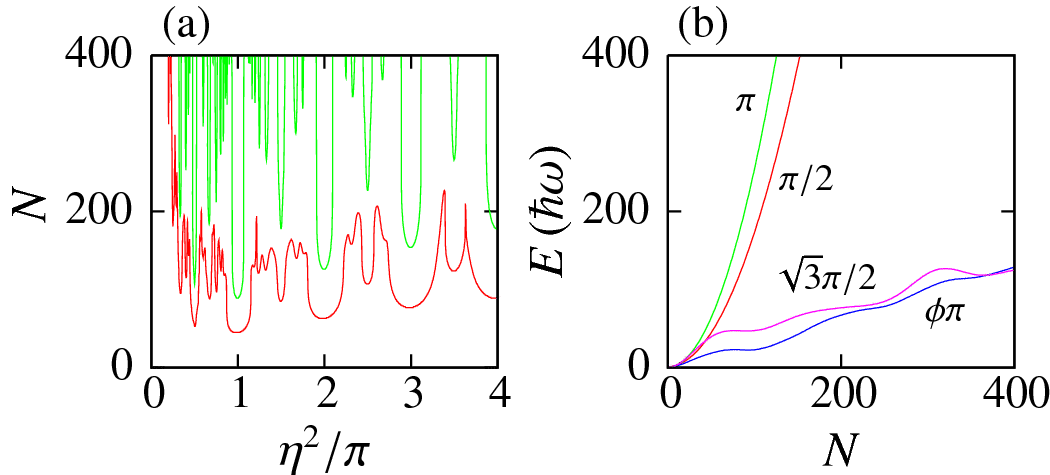}
\caption{(color online). Mean energies yielded by KHO evolution when $q=4$, $\kappa=-0.8$, and the initial condition is the harmonic oscillator ground state.  (a) Plots of the number of kicks required to reach a mean energy of $50\hbar$ (lower line) and $200\hbar \omega$ (upper line) as a function of $\eta^{2}$. (b) Mean energy as a function of $N$ for $\eta^{2} =\pi$, $\pi/2$, $\sqrt{3}\pi/2$,  and $\phi\pi$ [$\phi = (1+\sqrt{5})/2$ is the golden ratio].}
  \label{fig:resonances}
\end{figure}

In the KHO, the expansion through phase space changes from diffusive (energy growing linearly in time) to ballistic (energy growing quadratically in time) at the quantum resonance values.  In Fig.\ \ref{fig:resonances}(a) we have plotted the number of kicks required for the KHO to achieve particular mean energy values, as a function of $\eta^{2}$ when starting from an initial ground state.  This reveals distinct minima in the number of kicks required when $\eta^{2} = w\pi$, i.e., a quantum resonant value.  Other, less prominent minima are associated with higher-order resonances, i.e.,  when $\eta^{2}$ is a rational multiple of $\pi$, with the rate of energy absorption falling rapidly as the rational denominator increases.  Figure \ref{fig:resonances}(b) shows explicitly how the mean energy grows quadratically as a function of the number of kicks when $\eta^{2}=\pi$ or $\pi/2$ (ballistic expansion), quite distinct to the two cases we have shown when $\eta^{2}$ is an irrational multiple of $\pi$ (diffusive expansion).

\subsection{Hofstadter butterfly spectrum}
\begin{figure*}
\includegraphics[width=\textwidth]{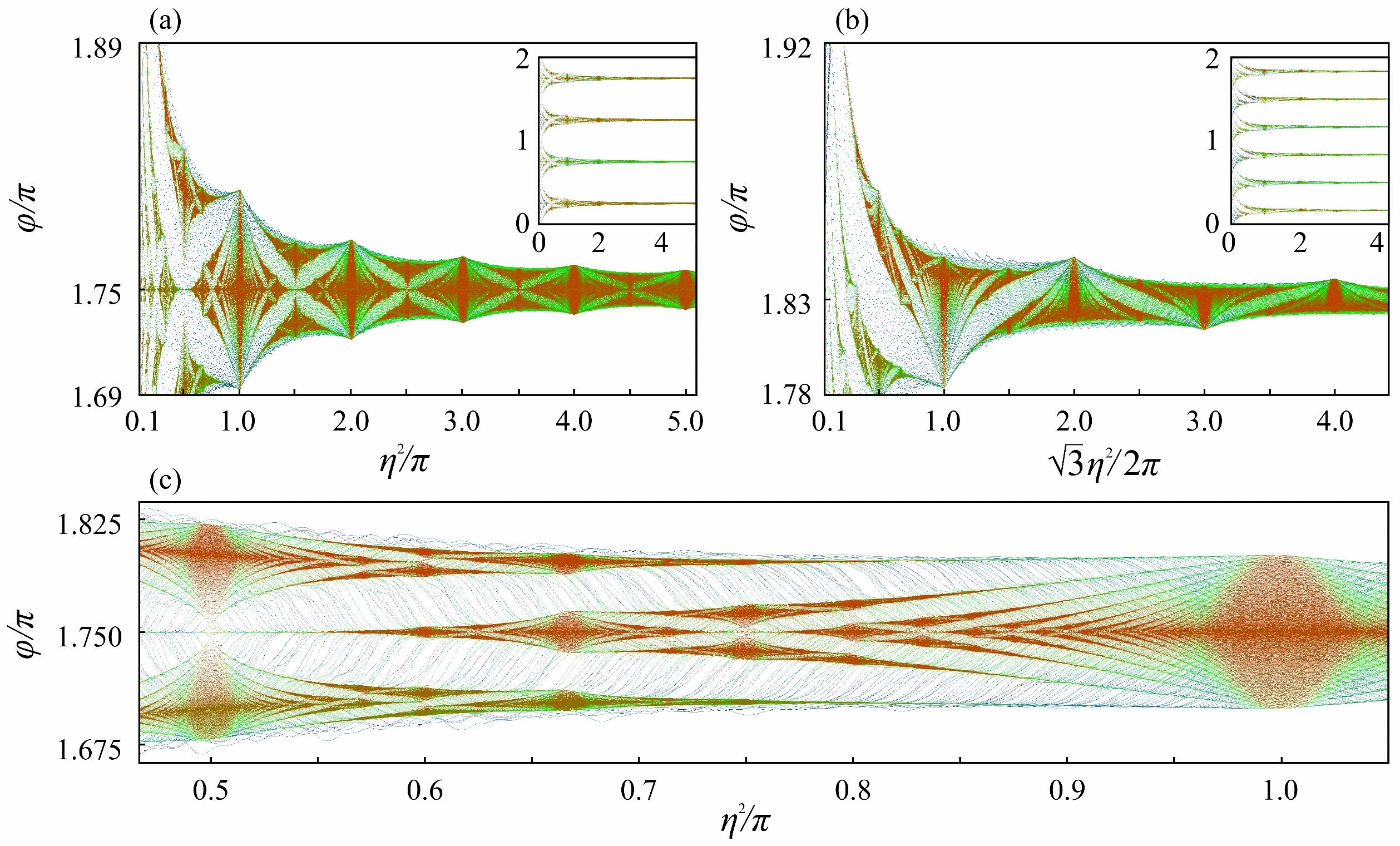}
\caption{(color online). Numerically determined quasienergy spectra of $\hat{F}$, computed in a truncated Fock basis of 2000 states and plotted against $\eta^2$. (a) and (b) show the overall structure (inset) and detail from the uppermost band in the cases $q=4$ and $q=6$, respectively.
A Hofstadter butterfly-type structure \cite{dana_dorofeev_2005_PRE, hofstadter_1976} is seen in each individual band, the cell boundaries (points at which the spectrum becomes continuous across the band) of which indicate the quantum resonant values of $\eta^{2}=w\pi$. 
One half of a cell for the case $q=4$ is shown in detail in (c), better illustrating the division of the band into $b$ continuous segments at rational fractions $a/b$ of the cell width. In each case $\kappa = -0.8$, and eigenvalues are colored according to the overlap of the associated eigenstate with the harmonic oscillator ground state, from blue ($\sim 10^{-23}$) to red ($\sim 10^{-1}$).}
\label{fig:floquet_spectrum}
\end{figure*}

The quantum resonances are also apparent in the quasienergy spectra of $\hat{F}$, i.e., the values $\varphi$ such that $e^{i\varphi}$ is an eigenvalue of $\hat{F}$. Taken as a function of $\eta^2$, the quasienergy spectrum of the KHO is a Hofstadter butterfly-type multifractal \cite{dana_dorofeev_2005_PRE, hofstadter_1976}, as shown in Fig.\ \ref{fig:floquet_spectrum}. The theory of butterfly-type spectra in kicked quantum maps has been extensively studied \cite{artuso_etal_1992,artuso_etal_1992_b,borgonovi_shepelyansky_1995,guarneri_1989,guarneri_1993,guarneri_borgonovi_1993,geisel_etal_1991}: In summary, a pure point spectral component is associated with localized behaviour, whilst singularly continuous and absolutely continuous spectral components are associated with diffusive and ballistic expansion respectively \cite{guarneri_1993,guarneri_1989}. For the KHO (when $q\in\{3,4,6\}$) this means that the quasienergy spectrum of $\hat{F}$ switches between being singularly continuous at irrational multiples, and absolutely continuous at rational multiples  of the primary quantum resonant values of $\eta^2$.

Kicked atom-optical systems have been proposed as candidates for direct observation of butterfly-type spectra \cite{wang_gong_2008}, and the atom-optical KHO considered here presents another system in which this might be realised. The KHO also offers the opportunity to directly explore the changes in the nature of the butterfly spectrum across the quantum resonance/antiresonance transition (that is, as a function of the relative alignment of the kicking and trapping potentials).  These changes have been explored theoretically for the case $q=4$  \cite{dana_dorofeev_2005_PRE}.

\subsection{Experimental realization}
To observe, for example, the principal quantum resonance in the case where $q=4$ requires a Lamb-Dicke parameter of $\eta \equiv K\sqrt{\hbar/2M\omega} = \sqrt{\pi}$, and hence a suitable combination of harmonic trapping angular frequency $\omega$ and wavevector $K$.  This could be realized in a linear ion trap with relatively weak axial confinement, perhaps in conjunction with a standing wave formed by two counterpropagating laser beams overlapping at an acute angle.  Typical experimental configurations involving laser cooled alkali atoms within a magnetic trap yield much larger values of $\eta$; a configuration equivalent to that used by Duffy \textit{et al}.\ would require a trap frequency of $4.8$ kHz \cite{duffy_etal_2004_PRA}.  Such frequencies are nevertheless within reach of recent generations of magnetic chip traps \cite{fortagh_and_zimmermann_2007}.  Alternatively, a potentially quite flexible set-up could be achieved by generating the kicking field with a CO$_2$ laser (wavelength $\lambda = 10.6 \mathrm{\mu m}$). Applying the kicking field directly along the axis of harmonic confinement, values of the Lamb-Dicke parameter around the principal quantum resonances (e.g., from $\eta^2 = \pi /2$ up to $\eta^2 = 2\pi$) could be reached using trap frequencies ranging from 8--36 Hz ($^{133}$Cs) up to 40--200 Hz ($^{23}$Na), and introducing the kicking field at an angle to the trap axis would allow these frequencies to be increased. Kicking strengths would be low because of the extreme detuning of the CO$_2$ laser, but the quantum resonances will still be manifest at any value of $\kappa$ after a sufficient number of kicks.

\section{Conclusions}
\label{sec:conclusions}
In this paper we have considered the characteristic quantum resonances occurring in an atom-optical $\delta$-kicked harmonic oscillator (KHO), a system in which typically alkali metal atom(s) or an alkaline earth ion are trapped in an approximately one-dimensional harmonic potential and periodically driven by pulses of an off-resonant laser standing wave. We have demonstrated that the atom-optical KHO displays quantum resonances dependent on the value of $\eta^2$, where $\eta$ is the Lamb-Dicke parameter, equal to the ratio of the width of the ground state of the harmonic trapping potential to the wavelength of the laser standing wave.

When the oscillator is kicked a rational number of times $r/q$ per oscillator period, with $q$ equal to $3$, $4$ or $6$, quantum resonances occur . At these values of $q$ the eigenstates of the Floquet operator $\hat{F}^q$ also possess a crystal phase space symmetry. We have shown that the quantum resonances occur at natural number multiples of a primary resonance value: $\eta^2 = \pi$ in the case $q=4$, and $\eta^2 = 2\pi/\sqrt{3}$ in the cases $q=3$ and $q=6$. At these values of $\eta^2$ all $q$ kicks within $\hat{F}^q$ commute. Further, secondary, quantum resonances occur for $\eta^2$ equal to non-integer rational multiples of the primary values; at these values of $\eta^2$ there is a partial commutativity between the kicks constituting $\hat{F}^q$.  We have demonstrated that the quantum resonances can be interpreted in terms of the relation between $\hat{F}^q$ and its own eigenstates; specifically, the quantum resonances occur when $\hat{F}^q$ can be decomposed in terms of a set of displacement operators which is equivalent to the set of crystal symmetries of the eigenstates.  We have also derived an expression for the time evolution of a coherent state in the cases $q=3,4,6$ by recasting the dynamics in terms of a mapping on a two-dimensional lattice of coefficients. At quantum resonance values of $\eta^2$ the evolution of these coefficients is greatly simplified; we have shown that in this case there is a direct expression for all the coefficients at a given step.

Finally, we have considered the manifestation of quantum resonances in an atom-optical KHO, in particular the ballistic increase in the mean energy associated with quantum resonances, and the quasienergy spectrum.  To this end we have also discussed experimentally reasonable parameter regimes in which to observe quantum resonant dynamics in the atom-optical $\delta$-kicked harmonic oscillator. 

\acknowledgements
We would like to thank C. S. Adams, K. R. Helmerson, I. G. Hughes, W. D. Phillips, and S. L. Rolston for stimulating and enlightening discussions, together with the UK EPSRC (Grant No.\ EP/D032970/1) and Durham University (T.P.B.) for support.

\appendix
\section{Quantum resonant values of $\eta^{2}$}
\label{app:derivation}

\begin{figure}
\includegraphics[width=\imwidth\columnwidth]{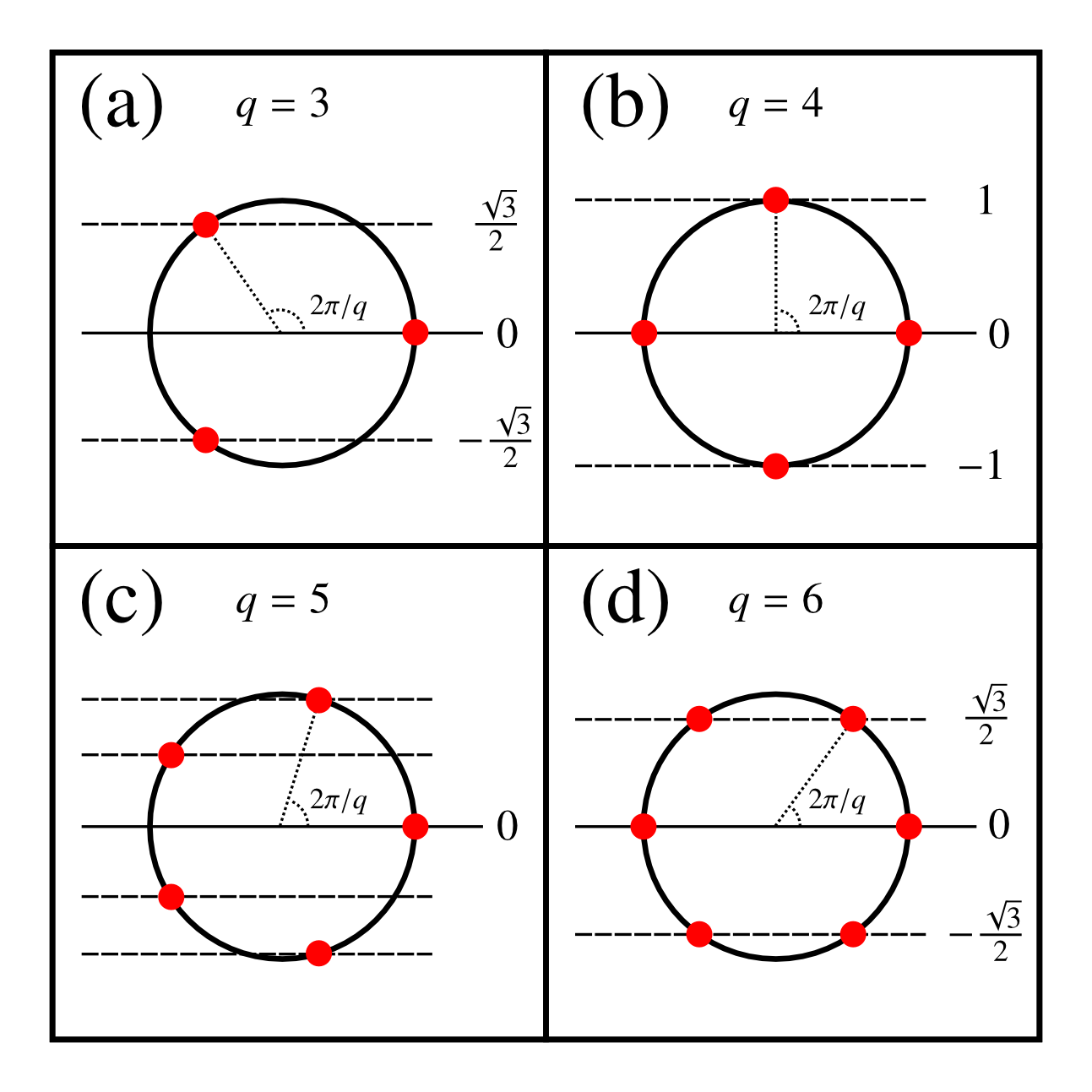}
\caption{(color online). Pictorial representation of the values of $\sin \left( 2\pi j / q \right)$ (the vertical position of points) in the cases $q=3$, $4$, $5$, and $6$.}
\label{fig:commutation_diagram}
\end{figure}

Because we are free to choose the sign of $w$ on the right side of \myref{eqn:master_resonance_condition} we can, without loss of generality, simplify that commutation condition to
\begin{equation}
\eta^2 \vert \sin \left( 2\pi j / q \right)\vert = w \pi; \, j \in [0,\ldots,q-1], \, w = 0,1,2,\ldots.
\label{eqn:new_master_resonance_condition}
\end{equation}
Quantum resonances occur when the condition \myref{eqn:new_master_resonance_condition} is satisfied for all possible $j$.

Ignoring the trivial cases $q=1$ and $q=2$, where $ \sin \left( 2\pi j / q \right) =0 $ for all possible $j$,  $ \vert \sin \left( 2\pi j / q \right) \vert$ takes either $(q-2)/2$ values other than zero (even $q$), or $(q-1)/2$ values other than zero (odd $q$). If we denote these nonzero values by $z_n$ [where $n$ ranges from $0$ to $(q-2)/2$ (even $q$), or from $0$ to $(q-1)/2$ (odd $q$)], the condition for quantum resonance becomes
\begin{equation}
 \eta^2 = w \pi / z_n, \quad \forall z_n.
\label{eqn:simple_condition}
\end{equation}

Because the $z_n$ are the moduli of sines equally distributed in angle, they cannot share a common factor other than unity. Consequently, condition \myref{eqn:simple_condition} can be satisfied for all $z_n$ only if all of the $z_n$ are equal. The cases of interest are enumerated pictorially in Figure \ref{fig:commutation_diagram}, where we see that: (a) $z_n=\sqrt{3}/2$ for $q=3$ ($\eta^2 = w2\pi/\sqrt{3}$ at quantum resonance); (b) $z_n=1$ for $q=4$ ($\eta^2 = w\pi$ at quantum resonance); (c) there is more than one value of $z_n$ for $q=5$ (there are no quantum resonances); (d) $z_n=\sqrt{3}/2$ for $q=6$ ($\eta^2 = w2\pi/\sqrt{3}$ at quantum resonance). Figure \ref{fig:commutation_diagram}(d) also illustrates the absence of quantum resonances for $q > 6$; the addition of another distinct point on the circumference of the circle shown necessitates at least a second value of $z_n$.

\section{Graf's addition theorem}
\label{app:graf}

\begin{figure}
\includegraphics[width=\imwidth\columnwidth]{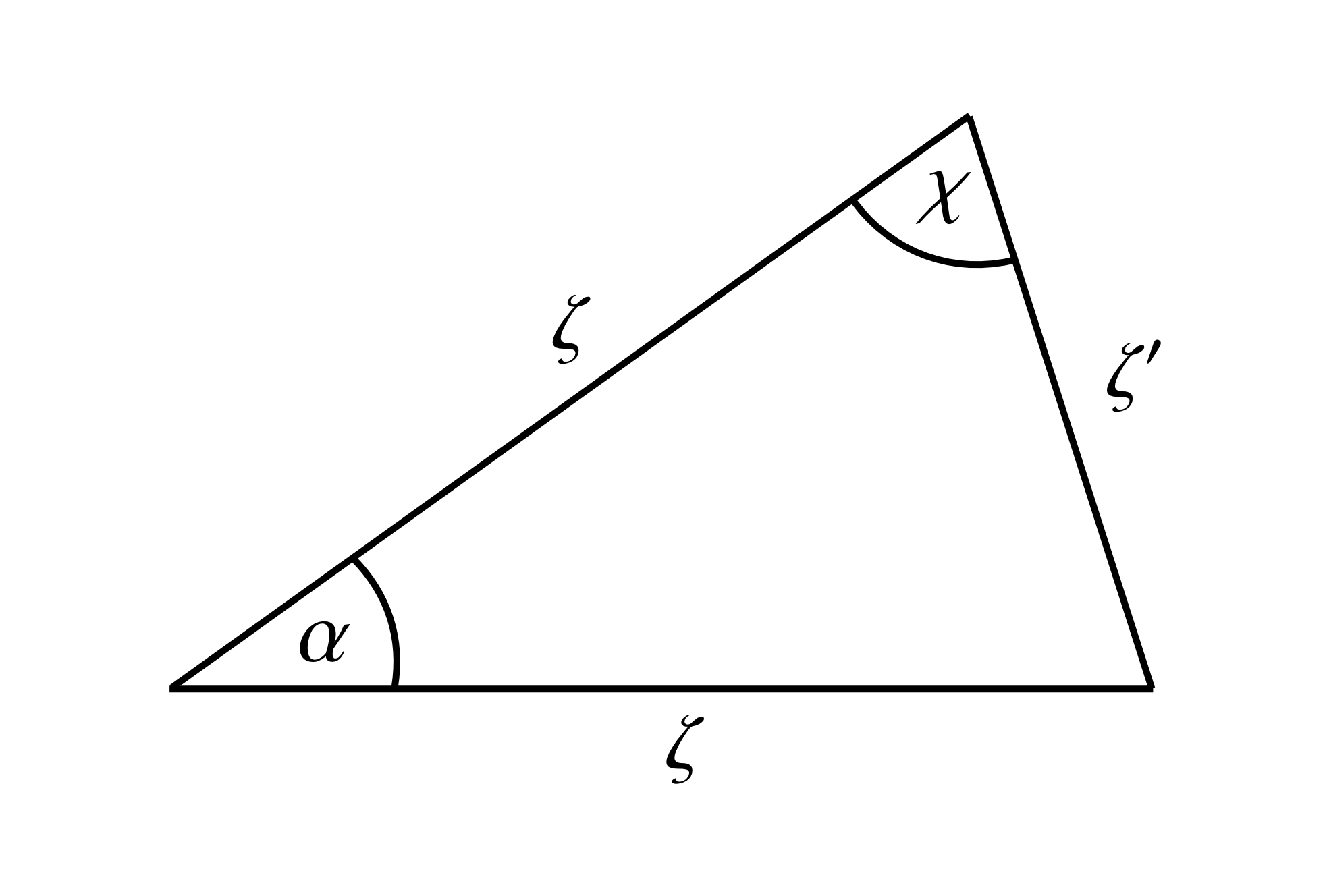}
\caption{Relation of variables appearing in Graf's addition theorem \myref{eqn:graf1} for two Bessel functions of integer order and equal, real, positive argument $\zeta$. Adapted from \cite{abramowitz_stegun}.}
\label{fig:graf_diagram}
\end{figure}

Graf's addition theorem for two Bessel functions of integer order and equal, real argument $\zeta$ is \cite{abramowitz_stegun}
\begin{equation}
 \sum_{k=-\infty}^{\infty} J_{n+k}(\zeta) J_{k}(\zeta) e^{ik\alpha} = J_{n}(\zeta^\prime) e^{in\chi},
\label{eqn:graf1}
\end{equation}
where $\zeta^\prime$ and $\chi$ are related to $\zeta$ and $\alpha$ by
\begin{align}
 \zeta^\prime &= \zeta \sqrt{2[1 - \cos(\alpha)]}, \\
 \zeta [1 - \cos(\alpha)] &= \zeta^\prime \cos(\chi), \\
 \zeta \sin(\alpha) &= \zeta^\prime \sin(\chi).
\end{align}
These relations are shown graphically for positive $\zeta$ in Fig.\ \ref{fig:graf_diagram}, for $0 \leq \alpha \leq \pi$.
In the quantum resonant case we consider in section \ref{subsubsec:q4}, $\alpha = \pi$. Consequently $\chi = 0$, and we obtain
\begin{equation}
 \sum_{k=-\infty}^{\infty} J_{n+k}(\zeta) J_{k}(\zeta) e^{ik\pi} = J_{n}(2\zeta).
\end{equation}

\section{Evolution of an initial coherent state when $q=6$}
\label{app:evo}
In the case $q=6$, and at quantum resonance $\eta^2=2\pi/\sqrt{3}$ the mapping \myref{eqn:coefficient_mapping} becomes
\begin{equation}
M^{[j+1]}_{m,n} = \sum_{k=-\infty}^{\infty} i^{k} J_{k}(\zeta) M^{[j]}_{m+n-k,-m} e^{-i k m \pi}.
\label{eqn:q=6mapping}
\end{equation}
As in the case $q=4$, we begin with the single coherent state initial condition
\begin{equation}
 M^{[0]}_{m,n} = \delta_{m,0} \delta_{n,0}.
\end{equation}
Replacing the indices in the correct fashion we get
\begin{equation}
 M^{[0]}_{m+n-k,-m} = \delta_{m+n,k} \delta_{m,0}
\end{equation}
and thus, substituting this into the mapping \myref{eqn:q=6mapping},
\begin{equation}
 M^{[1]}_{m,n} 
                            =  i^{m+n} J_{m+n}(\zeta) \delta_{m,0} e^{-i m(m+n) \pi}.
\label{eqn:q=6M1}
\end{equation}

Again, as in the case $q=4$, we observe the more general pattern of evolution by repeating this process until we obtain the coefficients $M^{[q]}_{m,n}$. We begin by rearranging \myref{eqn:q=6M1} to get
\begin{equation}
 M^{[1]}_{m+n-k,-m} = i^{n-k} J_{n-k}(\zeta) \delta_{m+n,k} e^{-i (m+n-k)(n-k) \pi}.
\end{equation}
Substituting this into the mapping \myref{eqn:q=6mapping} produces
\begin{equation}
M^{[2]}_{m,n} 
                                       = i^{m+n} J_{m+n}(\zeta) i^{m} J_{m}(\zeta) e^{-i m(m+n) \pi}, \label{eqn:q=6M2}
\end{equation}
where we have used the identity $i^{-n}J_{-n}(\zeta) = i^{n}J_{n}(\zeta)$. Rearranging \myref{eqn:q=6M2} we find that
\begin{equation}
M^{[2]}_{m+n-k,-m} = i^{n-k} J_{n-k}(\zeta) i^{m+n-k} J_{m+n-k}(\zeta) e^{-i (mn + n^2 + k^2 +mk) \pi},
\end{equation}
and hence
\begin{equation}
 M^{[3]}_{m,n} = \sum_{k=-\infty}^{\infty} i^{k} J_{k}(\zeta) i^{n-k} J_{n-k}(\zeta) i^{m+n-k} J_{m+n-k}(\zeta) e^{-i (mn +n^2 +k^2)\pi}.
\label{eqn:q=6M3}
\end{equation}

Graf's addition theorem (Appendix \ref{app:graf}) cannot be directly applied to \myref{eqn:q=6M3}, as there are three Bessel functions with $k$-dependent indices. However, we accept this summation as the final form of $M^{[3]}_{m,n}$ and continue with the process, using $k^\prime$ as the new index of summation. Rearranging \myref{eqn:q=6M3} we have
\begin{equation}
\begin{split}
M^{[3]}_{m+n-k^\prime,-m} =& \sum_{k=-\infty}^{\infty} i^{k} J_{k}(\zeta) i^{m+k} J_{m+k}(\zeta) i^{n-k-k^\prime} \\
& \times  J_{n-k-k^\prime}(\zeta) e^{i [(m+n-k^\prime)m -m^2 - k^2] \pi},
\end{split}
\end{equation}
which, when substituted into the mapping \myref{eqn:q=6mapping}, produces
\begin{equation}
\begin{split}
M^{[4]}_{m,n} = &  \sum_{k=-\infty}^{\infty} \sum_{k^\prime=-\infty}^{\infty} i^{k^\prime} J_{k^\prime}(\zeta)  i^{k} J_{k}(\zeta) i^{m+k} J_{m+k}(\zeta)  \\
 & \times i^{n-k-k^\prime} J_{n-k-k^\prime}(\zeta) e^{i (mn-k^2) \pi}, \\
= & \sum_{k=-\infty}^{\infty} i^{k} J_{k}(\zeta) i^{m+k} J_{m+k}(\zeta) i^{n-k} e^{i(mn - k^2)\pi} \\ 
 & \times \sum_{k^\prime=-\infty}^{\infty}  J_{k^\prime}(\zeta)  J_{n-k+k^\prime}(\zeta) e^{ik^\prime \pi}, \\
= & \sum_{k=-\infty}^{\infty} i^{k} J_{k}(\zeta) i^{m+k} J_{m+k}(\zeta) i^{n-k} J_{n-k}(2\zeta) e^{i(mn - k^2)\pi},
\end{split}
\end{equation}
where we have used Graf's theorem to eliminate the summation over $k^\prime$. Rearranging this we obtain
\begin{equation}
\begin{split}
M^{[4]}_{m+n-k^\prime,-m} = & \sum_{k=-\infty}^{\infty} i^{k} J_{k}(\zeta) i^{m+n-k^\prime+k} J_{m+n-k^\prime+k}(\zeta) \\
& \times i^{m+k} J_{m+k}(2\zeta) e^{-i[m(m+n-k^\prime) + k^2]\pi},
\end{split}
\end{equation}
and hence
\begin{equation}
\begin{split}
M^{[5]}_{m,n} = & \sum_{k=-\infty}^{\infty} \sum_{k^\prime=-\infty}^{\infty} i^{k^\prime} J_{k^\prime}(\zeta) i^{k} J_{k}(\zeta) i^{m+n-k^\prime+k} J_{m+n-k^\prime+k}(\zeta)  \\
& \times i^{m+k} J_{m+k}(2\zeta) e^{-i[m(m+n) + k^2]\pi}  \\
  =& \sum_{k=-\infty}^{\infty}i^{k} J_{k}(\zeta)  i^{m+k} J_{m+k}(2\zeta) i^{m+n+k}   e^{-i[m(m+n) + k^2]\pi} \\
 & \times  \sum_{k^\prime=-\infty}^{\infty} J_{k^\prime}(\zeta)  J_{m+n+k^\prime+k}(\zeta) e^{ik^\prime \pi}\\
  =& \sum_{k=-\infty}^{\infty}i^{k} J_{k}(\zeta)  i^{m+k} J_{m+k}(2\zeta) i^{m+n+k} J_{m+n+k}(2\zeta)\\
 & \times   e^{-i[m(m+n) + k^2]\pi}.
\end{split}
\end{equation}
Repeating the process one final time we have
\begin{equation}
\begin{split}
M^{[5]}_{m+n-k^\prime,-m} = & \sum_{k=-\infty}^{\infty}i^{k} J_{k}(\zeta)  i^{m+n-k^\prime+k} J_{m+n-k^\prime+k}(2\zeta) \nonumber \\
 & \times i^{n-k^\prime+k} J_{n-k^\prime+k}(2\zeta)   e^{-i(mn + n^2 +{k^\prime}^2 + k^2 -mk^\prime)\pi},
\end{split}
\end{equation}
and hence
\begin{equation}
\begin{split}
M^{[6]}_{m,n} =&  \sum_{k=-\infty}^{\infty} \sum_{k^\prime=-\infty}^{\infty} i^{k^\prime} J_{k^\prime}(\zeta)i^{k} J_{k}(\zeta)  i^{m+n-k^\prime+k} J_{m+n-k^\prime+k}(2\zeta) \\
& \times i^{n-k^\prime+k} J_{n-k^\prime+k}(2\zeta)   e^{-i(mn + n^2 +{k^\prime}^2 + k^2)\pi}. \label{eqn:q=6M6_1}
\end{split}
\end{equation}

If we define $k=k^\prime +\Delta$, then \myref{eqn:q=6M6_1} takes the form
\begin{equation}
\begin{split}
M^{[6]}_{m,n} =&  \sum_{\Delta=-\infty}^{\infty}  i^{m+n+\Delta} J_{m+n+\Delta}(2\zeta) i^{n+\Delta} J_{n+\Delta}(2\zeta) \\ & \times  i^\Delta  e^{-i(mn + n^2  + \Delta^2)\pi}\sum_{k^\prime=-\infty}^{\infty} J_{k^\prime}(\zeta) J_{k^\prime+\Delta} (\zeta) e^{i k^\prime \pi}. \label{eqn:q=6M6_2}
\end{split}
\end{equation}
By applying Graf's theorem to the summation over $k^\prime$ in \myref{eqn:q=6M6_2}, and then applying the transformation $\Delta \rightarrow -\Delta$ we obtain
\begin{equation}
 \begin{split}
M^{[6]}_{m,n} = & \sum_{\Delta=-\infty}^{\infty} i^\Delta J_{\Delta}(2\zeta) i^{n-\Delta} J_{n-\Delta}(2\zeta)  \\ & \times i^{m+n-\Delta} J_{m+n-\Delta}(2\zeta)   e^{-i(mn + n^2  + \Delta^2)\pi},
\end{split}
\end{equation}
that is, an analogue of $M^{[3]}_{m,n}$ with $\zeta$ replaced by $2\zeta$ and $k$ replaced by $\Delta$. As in the case $q=4$, the evolution from this point on will continue cyclically.


\end{document}